\documentclass[10pt]{iopart}
\usepackage{graphicx}
\usepackage{hyperref}

\newcommand \co  {$^{13}$CO \ }
\newcommand \kms {km s$^{-1} \ $}
\newcommand \as  {$^{\prime\prime}~$}
\newcommand \am  {$^{\prime}~$}

\hypersetup{colorlinks,citecolor=blue,linkcolor=blue,urlcolor=blue}

\begin{document}
\title[The kinematic distances of SNR G16.7+0.1 and G15.9+0.2 by analyzing the HI absorption spectra]{The kinematic distances of SNR G16.7+0.1 and G15.9+0.2 by analyzing HI absorption spectra}

\author{W.W. Tian$^{1,2}$, H. Zhu$^{1}$, M.F. Zhang$^{1,2}$, H.K. Chang$^{4,5}$,  S.S. Shan$^{1,2}$ \& D.A. Leahy$^{3}$}
\address{$^1$Key Laboratory of Optical Astronomy, National Astronomical Observatories, Chinese Academy of Sciences,
Beijing 100012, China\\
$^2$University of Chinese Academy of Sciences, 19A Yuquan Road, Shijingshan District, Beijing 100049, China\\
$^3$Department of Physics $\&$ Astronomy, University of Calgary, Calgary, Alberta T2N 1N4, Canada\\
$^4$Institute of Astronomy, National Tsing Hua University, Hsinchu 30013, Taiwan\\
$^5$Department of Physics, National Tsing Hua University, Hsinchu 30013, Taiwan\\}
\ead{tww@bao.ac.cn, zhuhui@bao.ac.cn}
\vspace{10pt}
\begin{indented}
\item[]August 2017
\end{indented}

\begin{abstract}
We build HI absorption spectra towards Supernova Remnant (SNRs) G16.7+0.1 and G15.9+0.2 using the \textit{THOR} survey data. With the absorption spectra, we give a new distance range of 7 to 16 kpc for G15.9+0.2. We also resolve the near/far-side distance ambiguity of G16,7+0.1 and confirm its kinematic distance of about 14 kpc. In addition, we analyze the CO (J=3-2) spectra towards G16.7+0.1 and find obvious CO emission at the 20 \kms OH 1720 MHz maser site. This supports Reynoso and Mangum (2000)'s suggestions that the velocity difference between the maser and southern molecular cloud is caused by the shock acceleration. We discuss the impact of the distances on other physical parameters of the two SNRs.
\end{abstract}
\vspace{2pc}
\noindent{\it Keywords}:ISM: individual objects: SNR G16.7+0.1,SNR G15.9+0.2 -- ISM: supernova remnants

\section{Introduction}
Supernova remnants (SNRs) play an important role in understanding supernova explosion mechanisms, the acceleration of Galactic cosmic rays and the chemical enrichment of the interstellar medium (ISM). SNRs' distances are key to study SNRs themselves and their environment. However, it is always a challenging job to measure the distance of Galactic SNRs. Up to now, there is a total of 295 SNRs in Green SNR Catalog\footnote{http://www.mrao.cam.ac.uk/surveys/snrs/}, for which the distances of just about 97 SNRs have been given.\\

Many methods have been used to determine the distances of SNRs, e.g. the empirical surface brightness ($\Sigma$) - diameter(D) relation which might include error of about 40\% or up to an order of magnitude \cite{Case1998,Green2005}; the widely-used kinematic distances based on an axisymmetric rotation curve model of the Milk Way \cite{Caswell1975}; the expansion distances by measuring the proper motions and radial velocities \cite{Green1984};  and the distances from the extinction($\rm A_V$)-distance relation, when the hydrogen column density ($\rm N_H$) or $\rm A_V$ towards the remnant is known \cite{shan2018},\cite{Zhu2015},\cite{Zhu2017}.\\

21 cm HI observations have traditionally been used for kinematic distances of SNRs since 1960 because HI atomic clouds are broadly distributed throughout the Galactic plane \cite{Dickey1990} and most of SNRs have radio emissions. 57 of the 97 distance-known SNRs were measured based on 21 cm HI observation. There are some limits for the method: (1) HI observation can generally provide a distance estimate to precision of a few hundred parsecs because of the inaccuracies in circular rotation curve models (e.g. \cite{Reid2014}, \cite{Brand1993}). However, if the HI cloud is located at region where the velocity field has significant deviation from the circular rotation model, the uncertainty in kinematic distance could be as large as a few kpc (e.g., \cite{Xu2006}). (2) The near/far kinematic distance ambiguity exists in the inner Galaxy caused by the kinematic model. (3) And finally, it is always a challenge to construct a reliable HI absorption spectrum to an SNR, especially for a faint and extended SNR due to spatial variation in the background HI emission over the SNR region.\\

For single dish or single-dish + interferometry HI absorption observation, the limitation of (3) becomes very important, because it relies on the assumption that the HI distribution along the target direction and the selected background direction are the same. We started systematic distance measurements to Galactic SNRs by improved methods to build 21 cm HI absorption spectrum against a background extended radio source \cite{Tian2007, Leahy2010}. In comparison with the traditional method-selecting background beside the source, we select the background surrounding the source directly to minimize the possibility of false absorption spectrum (see Fig.2 of \cite{Leahy2010}, and also see Fig.1 of \cite{Leahy2008} showing "on" and "off" regions). In addition, the CO spectrum toward a target source and the HI absorption spectra of other nearby bright continuum sources are also used when they exist to understand the absorption spectrum of an SNR better. So far we have obtained new or revised distances for more than 20 SNRs \cite{Tian2008a, Zhu2014, Ranasinghe2017}. Several distances are significantly revised, e.g. the distance of the SNR Kes 75/PSR J1846-0258 system changed from about 21 kpc to about 6.3 kpc \cite{Leahy2008}; the TeV SNR G349.7+0.2 has a revised distance from 22 kpc to 11.5 kpc \cite{Tian2014}. Our distance revision to SNR W51C is enhanced by newly discovering its northeast edge and the MHD simulation \cite{Zhang2017}.\\

If the HI absorption spectra are from interferometric observation, the uncertainty caused by (3) is significantly reduced. The HI emission will be filtered out when its angular size is larger than ${\lambda}/{B}_{min}$, where ${\lambda}$ is the wavelength and ${B}_{min}$ represents the minimum baseline length (\cite{Dickey1990}, \cite{Kolpak2003}). Take \textit{THOR} survey for example. The minimum baseline length is 35 m. At 1.4 GHz, the maximum angular scale that \textit{THOR} survey sensitive to is about 16\am. That means, if the angular scale of the target source is smaller than 16\am, the HI absorption is measured directly with no need of background HI emission subtraction.\\

We here study the distances of SNR G16.7+0.1 and G15.9+0.2 by employing the recently released \textit{THOR} data (The HI OH Recombination line survey of Milky Way \cite{Beuther2016}). SNR G16.7+0.1 is a composite SNR, detected at radio and X-ray wavelengths. The SNR is associated with PWN G16.73+0.08 and interacting with a nearby molecular cloud (MC) \cite{Green1997, Reynoso2000,Chang2018}. SNR G15.9+0.2 discovered by Clark et al. (1975) \cite{1975AuJPA..37....1C} has a spectral index of 0.56$^{+0.03}_{-0.03}$ \cite{Sun2011}, an age of (2900$\pm$200)-(5700$\pm$300)yr, a size of 6.5\am $\times$ 4.7\am and an brightness enhancement at the eastern border \cite{Caswell1982,Sasaki2018}. There are no reliable distances to the two SNRs in previous studies (see Green catalog: https://www.mrao.cam.ac.uk/surveys/snrs/).

\section{Data and Methods for  HI Absorption Spectra }
We use the data of \textit{THOR} DR1, which consists of 1.4 GHz continuum maps and HI, OH spectral lines and radio recombination lines data, covering the region $15^{\circ}<l<67^{\circ}$, $-1^{\circ}<b<1^{\circ}$ \cite{Beuther2016}.
The data has an angular resolution of about 20\as.
The velocity resolutions of both HI and OH spectral lines data are 1.5 \kms,
with velocity coverage from -127 to +171 \kms and -58.5 to +135 \kms, respectively.
The rms per channel after smoothing to a uniform beam of 21\as $\times$ 21\as is 3 K.\\

We also use \textit{GRS} archive data \cite{Jackson2006} to extract the \co spectra.
This survey provides data for the region $14^{\circ}<l<18^{\circ}$, and its angular resolution is 46\as.
Its velocity range and resolution are from -5 to 135 \kms and 0.21 \kms.
Its typical rms is 0.13 K per channel. The CO (J=3-2) data from \textit{COHRS} is also used to trace high density clouds\cite{Dempsey2013}.
The spatial resolution and mean rms of the data are 16\as and $\sim$1 K respectively.\\

\begin{figure}
    \centering
    \includegraphics[width=0.8\textwidth]{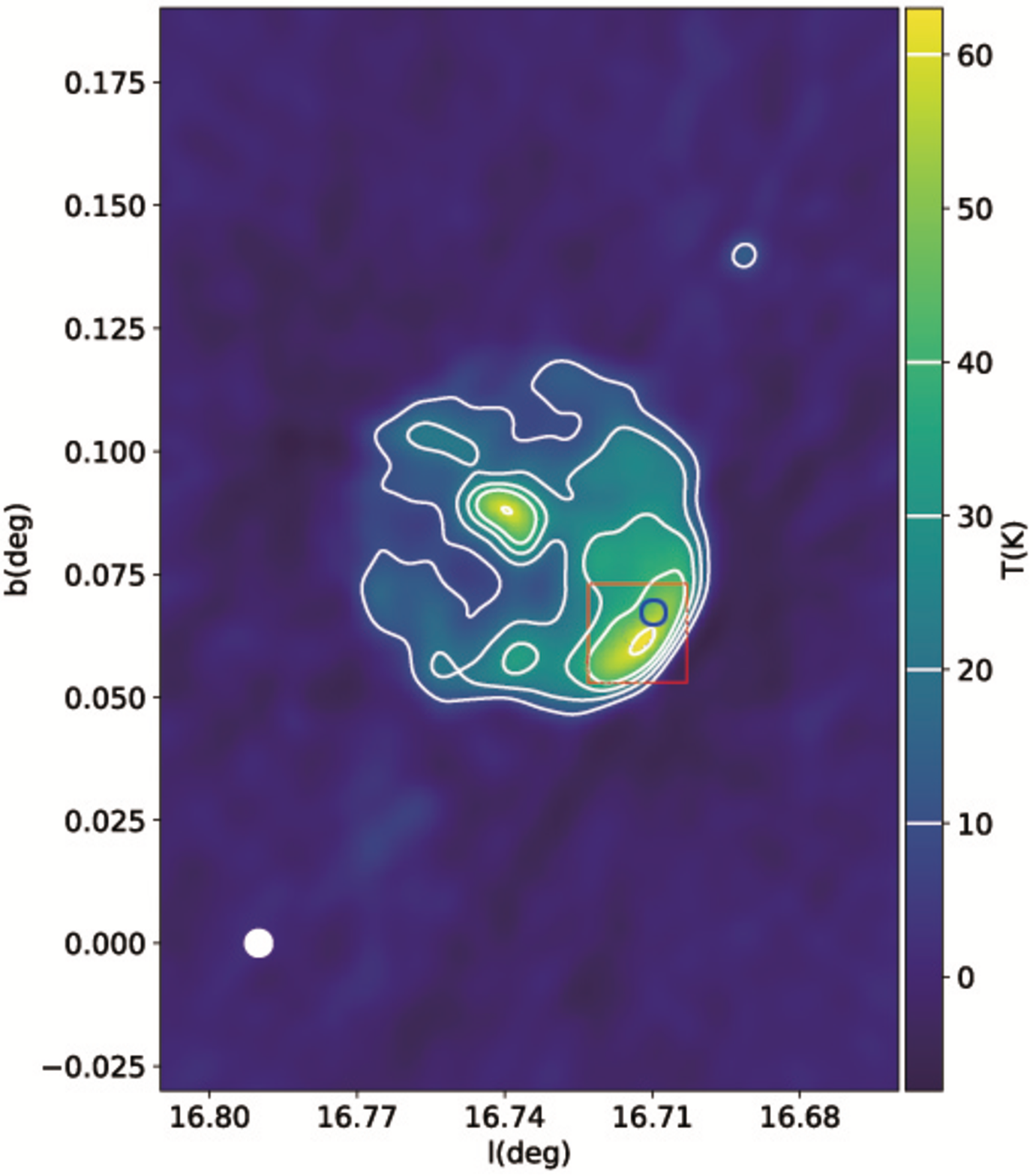}
\includegraphics[width=0.8\textwidth]{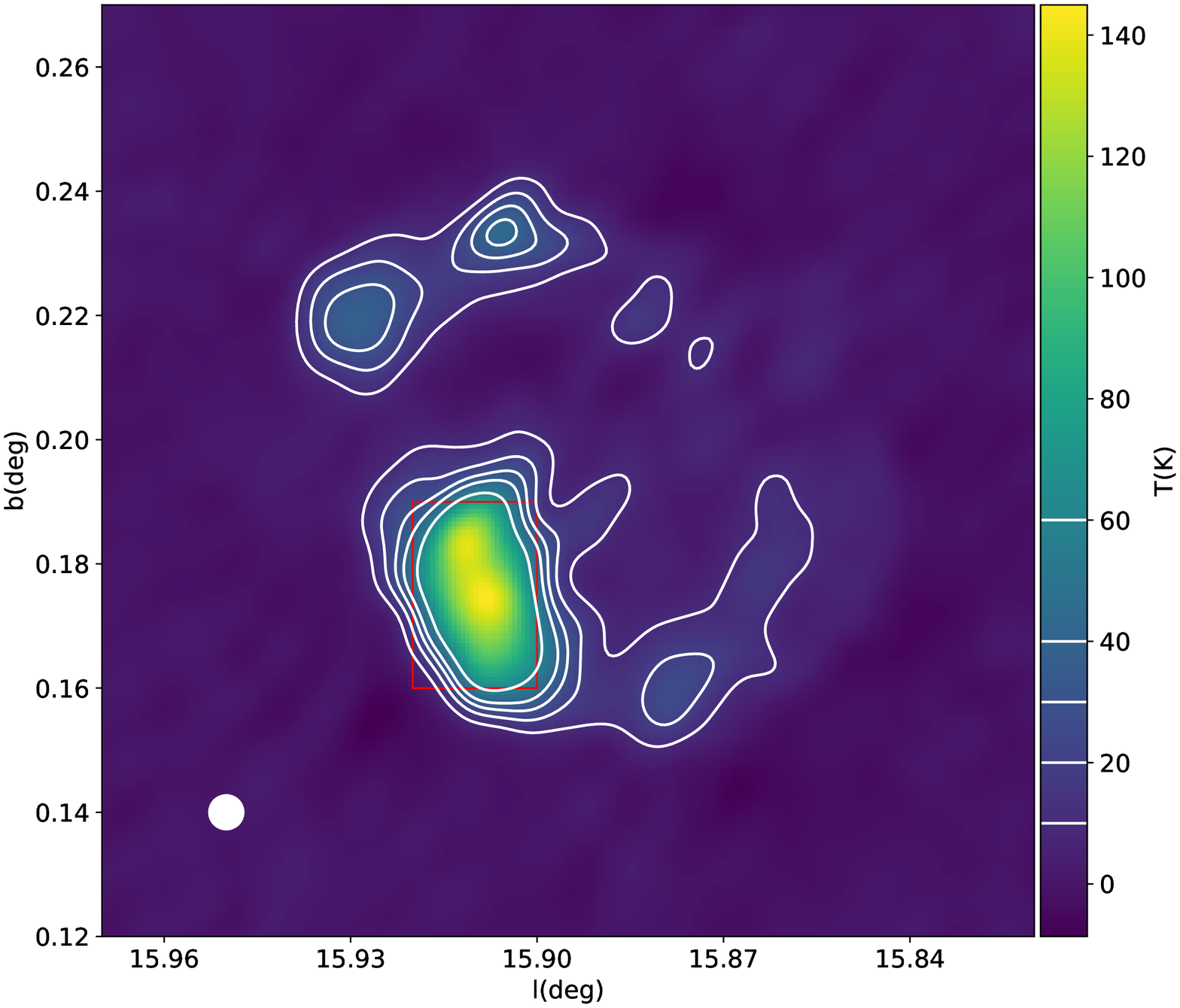}
        \caption{1.4 GHz continuum maps for G16.7+0.1 and G15.9+0.2. The red box in the top panel shows 'on' region of G16.7+0.1.  The 1720 MHz OH maser at 20 \kms is labelled by a blue circle. The red box in the bottom panel shows 'on' region of G15.9+0.2. The contours of the maps are shown as white lines (10k, 20k, 30k, 40k, 50k) and the beam size (20\as) of the survey is shown as a white circle at the lower left of the two images.}
\label{fig:on_off}
\end{figure}

Fig.~\ref{fig:on_off} shows the 1.4 GHz continuum images of G15.9+0.2 and G16.7+0.1 from \textit{THOR}.
We plot the boxes in Fig.~\ref{fig:on_off} to extract the HI absorption and the \co emission spectra (see Fig.~\ref{fig:spectra}).
Based on the data of \textit{THOR}, we also analyze the OH lines at 1720, 1665, 1612 MHz and extract the spectrum of 1720 MHz OH maser towards the 'on' region of G16.7+0.1.

\section{Results}
\begin{figure}
    \centering
    \includegraphics[width=0.8\textwidth]{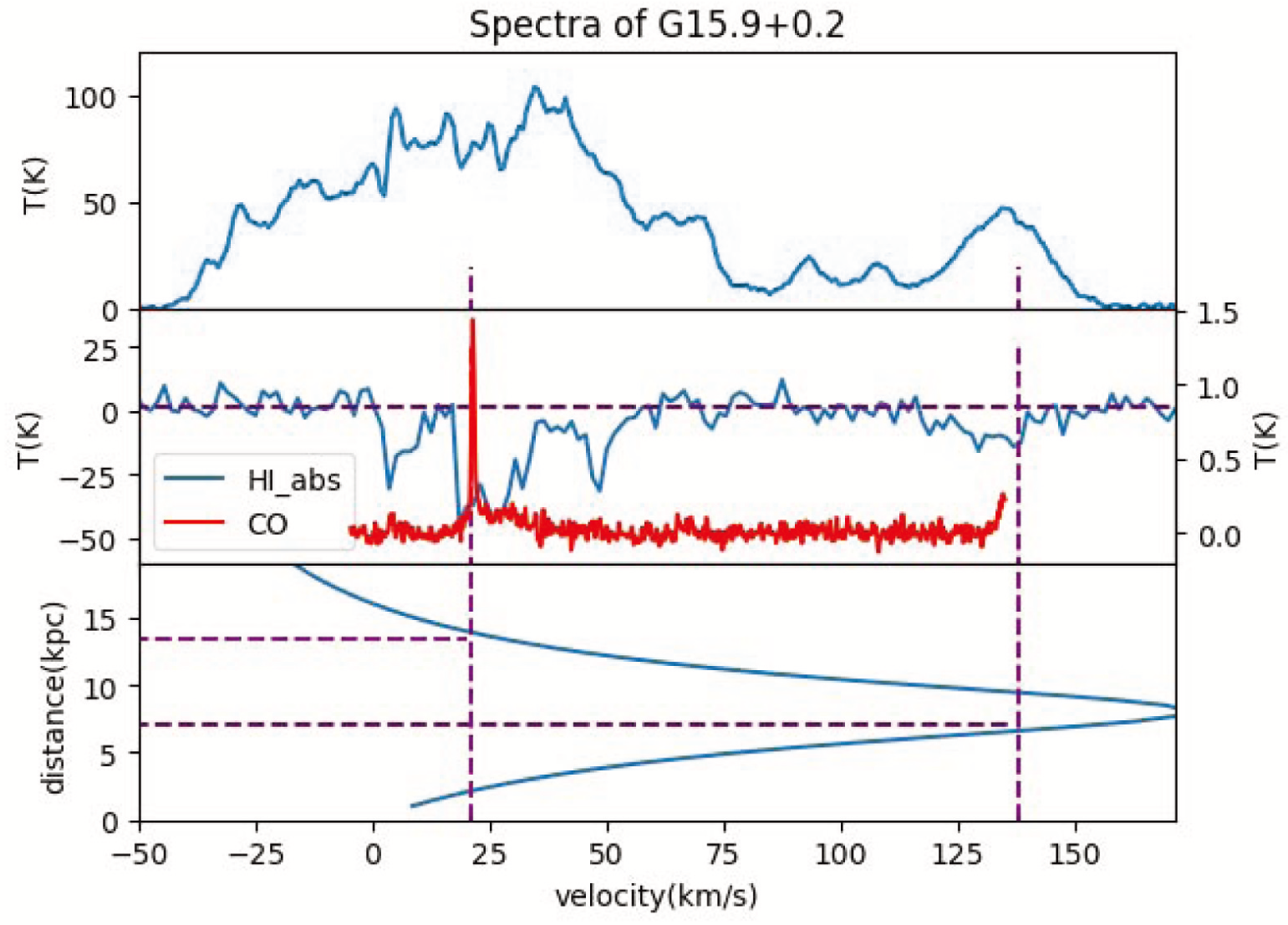}
    \caption{HI emission and absorption spectra of G15.9+0.2 and the distance-velocity relation towards this remnant. The blue line in the top panel is the HI emission spectrum from \textit{SGPS}\cite{Mcclure2005}. The blue and red lines in middle panel are the HI absorption and \co emission spectra, respectively, corresponding to the left and right y-axis label. The blue line in bottom panel shows the distance-velocity relation. The purple vertical lines indicate the velocities of 20 and 138 \kms. The purple horizontal lines in all panels present distances at 7 and 14 kpc.}
\label{fig:spectra}
\end{figure}

Fig.~\ref{fig:spectra} shows the HI absorption spectrum, \co emission spectrum, and the distance-velocity relation towards SNR G15.9+0.2. The distance-velocity relation (for details see Ranasinghe and Leahy 2017, Fig.2 and Section 3.2) is based on the circular Galactic rotation model (V$_{0}$ = 240 \kms, R$_{0}$ = 8.34 kpc) \cite{Reid2014}. The maximum velocity of HI absorption is at about 138 \kms, hinting that G15.9+0.2 is beyond about 7 kpc, the near-side distance of 138 \kms. We do not detect any absorption features at negative velocity, which indicates that G15.9+0.2 is within the solar circle (see Fig.~\ref{fig:4}) and has an upper limit of about 16 kpc.\\

\begin{figure}
    \centering
    \includegraphics[width=0.8\textwidth]{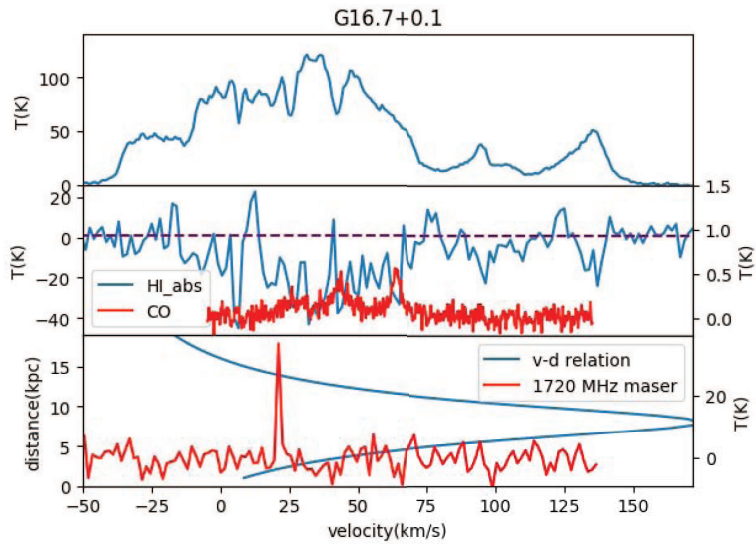}
    \caption{HI emission and absorption spectra of G16.7+0.1 and distance-velocity relation towards this remnant. The blue line in the top panel is the HI emission spectrum from \textit{SGPS}\cite{Mcclure2005}. In the middle panel, The HI absorption spectrum is in blue, and \co emission is in red. The blue and red lines in the bottom panel show the distance-velocity relation and 1720 MHz emission.}
\label{fig:spectra1}
\end{figure}

\begin{figure}
    \centering
    \includegraphics[width=0.8\textwidth]{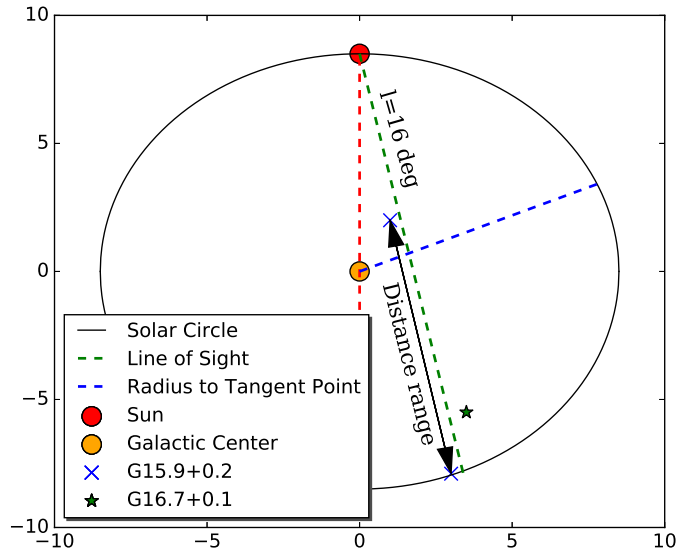}
    \caption{Schematic view of the Galactic plane from the Galactic north pole. Axes are labelled by x and y coordinate distances from the Galactic center in units of kpc. The black circle is the solar circle. The locations of G15.9+0.2 and G16.7+0.1 are indicated by the line with arrows and the star, respectively.}
\label{fig:4}
\end{figure}

In Fig.~\ref{fig:spectra1}, we show the HI absorption, \co emission and 1720 MHz emission spectra for G16.7+0.1. A bright single 1720 MHz OH maser line can be seen at 20 \kms. There are multiple HI absorption features at about 5, 25, 42, 65, 96 and 138 \kms, and also \co emission features at 25, 42 and 65 \kms simultaneously.
Since G16.7+0.1 shows absorption up to the tangent point, therefore, it must be located between the tangent point and the far side of the solar circle.\\

\section{Discussion}

\subsection{SNR G16.7+0.1}
Green et al. (1997) \cite{Green1997} detected the 1720 MHz OH maser at 20 \kms in the SNR G16.7+0.1, and suggested either a near-side distance of 2.2 kpc or a far-side distance of about 14.1 kpc.
Using both VLA and Green Bank 100 m telescope, Hewitt et al. (2008) \cite{Hewitt2008} identified the maser and showed the spectra of all four OH transitions towards G16.7+0.1.\\

We show the maser's integrated line intensity in Fig.~\ref{fig:spectra1} (the bottom panel). The strong 1720 MHz OH maser emission can be seen at 20 \kms which is a clear evidence that SNR G16.7+0.1 is interacting with a 20 \kms MC at $\sim$2 kpc or $\sim$14 kpc.
Since HI absorption features can be seen from 25 \kms to 138 \kms, we confirm that the remnant is located at the far-side distance of about 14 kpc. This distance is also supported by the large X-ray absorbing column density of the PWN in G16.7+0.1 \cite{Helfand2003}. In Table 1, we summary the key HI absorption and CO emission features to determine the distance of about 14 kpc. Note that the kinematic distance from the maser velocity might suffer the uncertainties from the considerable velocity difference between the maser and molecular cloud.

Previous CO (J=1-0, J=2-1) observations showed a small MC at $\sim$25 \kms at the maser site \cite{Reynoso2000, Kilpatrick2016} (hereafter the southern cloud). The broadened CO (J=2-1) indicates a potential interaction between G16.7 and the southern cloud. A weak wing centered at 19.3 \kms with a 6.5 \kms FWHM, has been found towards the southern cloud indicating that the maser is associated with the southern cloud and the $\sim$5 \kms velocity gap may be caused by the shock acceleration. However, the number density of the southern cloud is only 260 cm$^{-3}$ based on the CO (J=1-0) emission \cite{Reynoso2000} which is much lower than the canonical H$_2$ density, 10$^5$ cm$^{-3}$, required to generate the 1720 MHz OH maser \cite{Lockett1999}. Since CO (J=1-0) line emission could be optical thick if the MC density is too high, higher density tracers are needed to resolve the above inconsistence.\\

\begin{figure*}
\centerline{\includegraphics[width=0.33\textwidth]{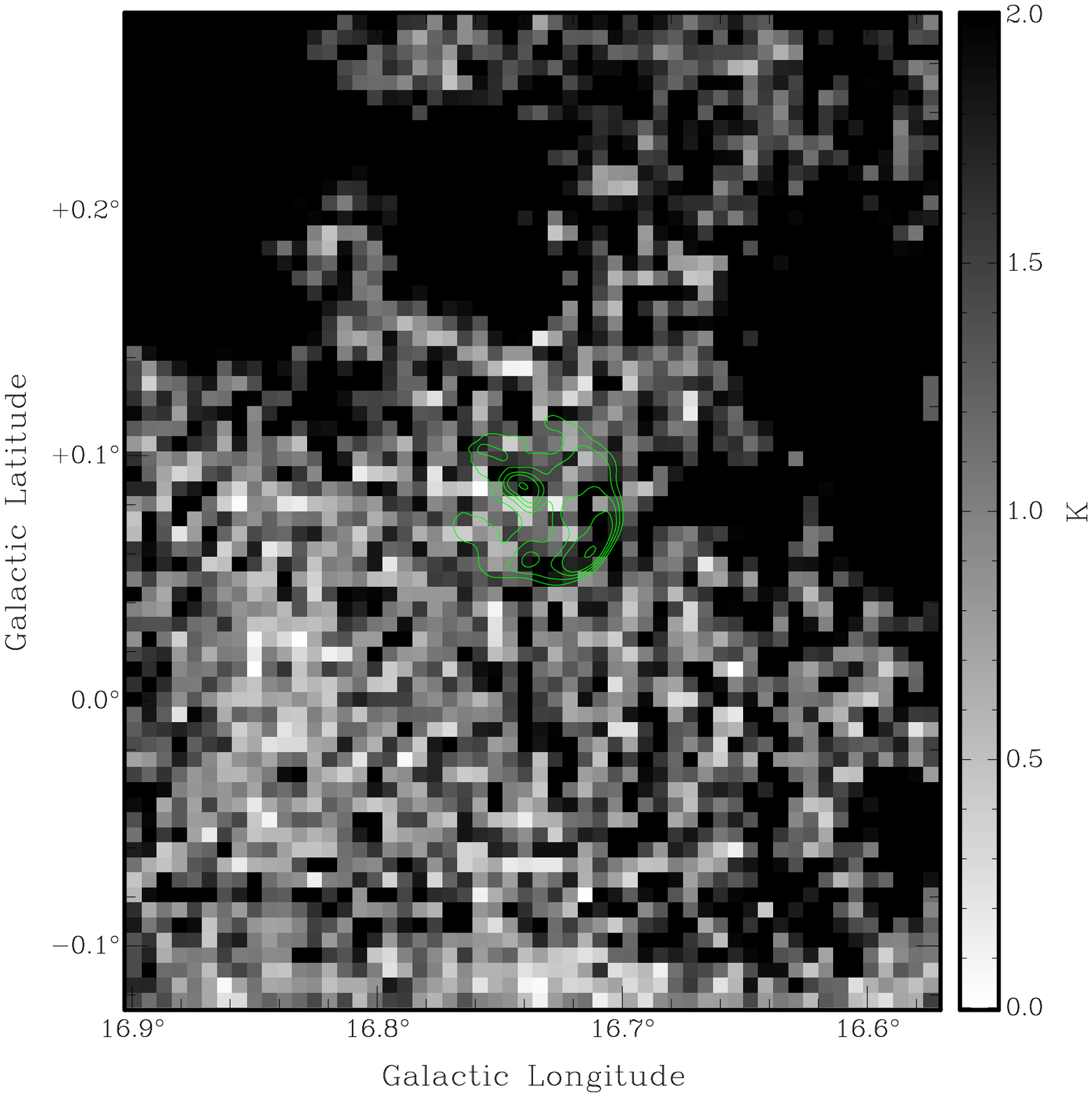}
\includegraphics[width=0.33\textwidth]{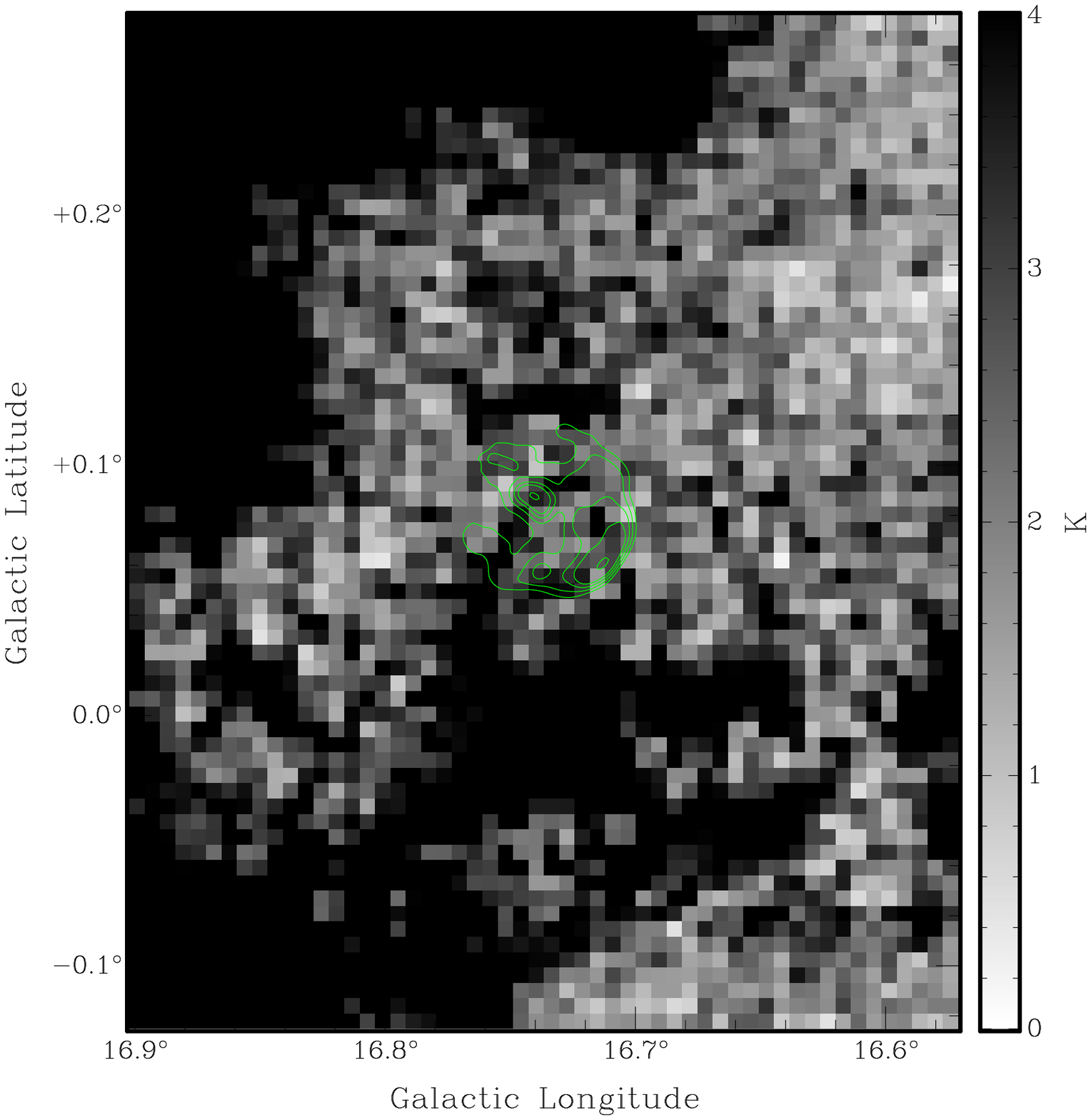}
\includegraphics[width=0.33\textwidth]{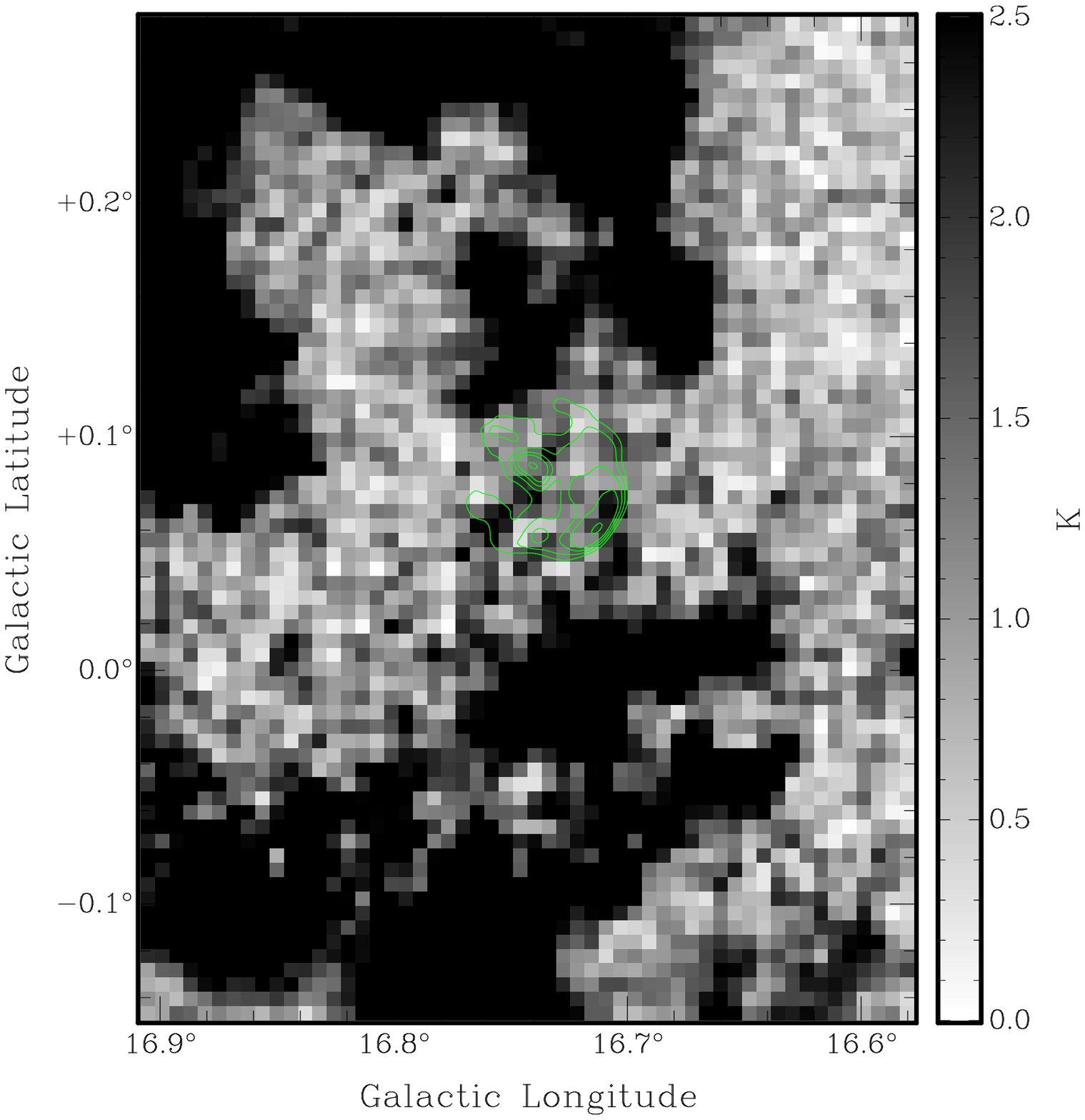}}
\caption{G16.7+0.1: The \co channel map integrated over 18-21, 22-24 and 25-26 \kms. The contours are the same as Fig.1.}
\label{fig:13co}
\end{figure*}

$^{13}$CO (J=1-0) and CO (J=3-2) line are good high density tracers. In Fig.~\ref{fig:13co}, we show the $^{13}$CO (J=1-0) emission maps integrated from 18 \kms to 21 \kms, 22 \kms to 24 \kms and 25 \kms to 26 \kms. The 25 \kms southern cloud can be seen clearly in the 25-26 \kms map. The location of the cloud in our map is consistent with Reynoso \& Mangum's Fig.7 \cite{Reynoso2000}. The difference is that the $^{12}$CO emission is stronger and more extended than the $^{13}$CO emission. The peak brightness temperature of the $^{13}$CO (J=1-0) emission is about 0.5 K (see Fig.~\ref{fig:spectra1}). There is no obvious emission in the 18-21 \kms and 22-24 \kms maps.\\

\begin{figure*}
\centerline{\includegraphics[width=0.33\textwidth]{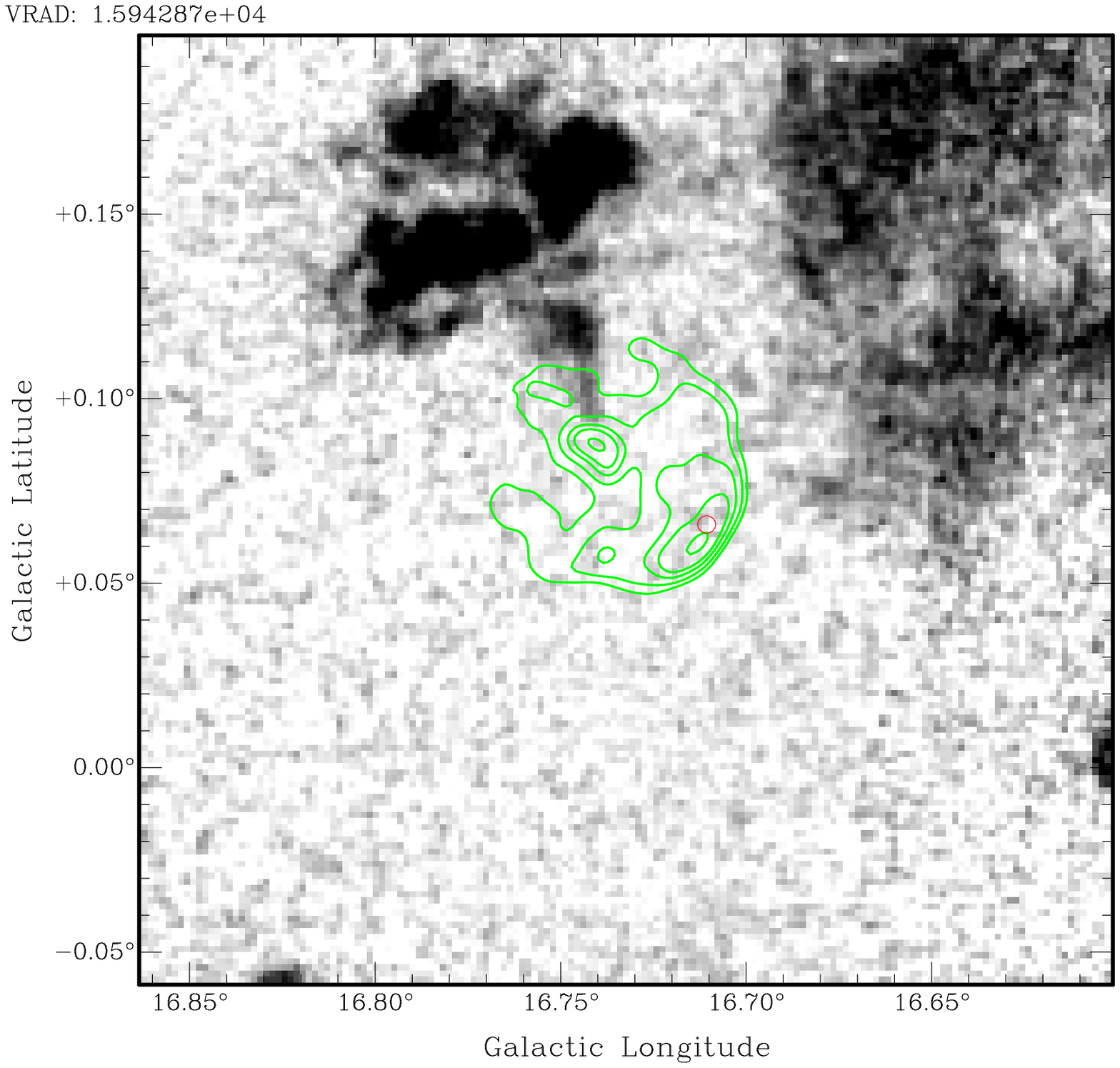}\includegraphics[width=0.33\textwidth]{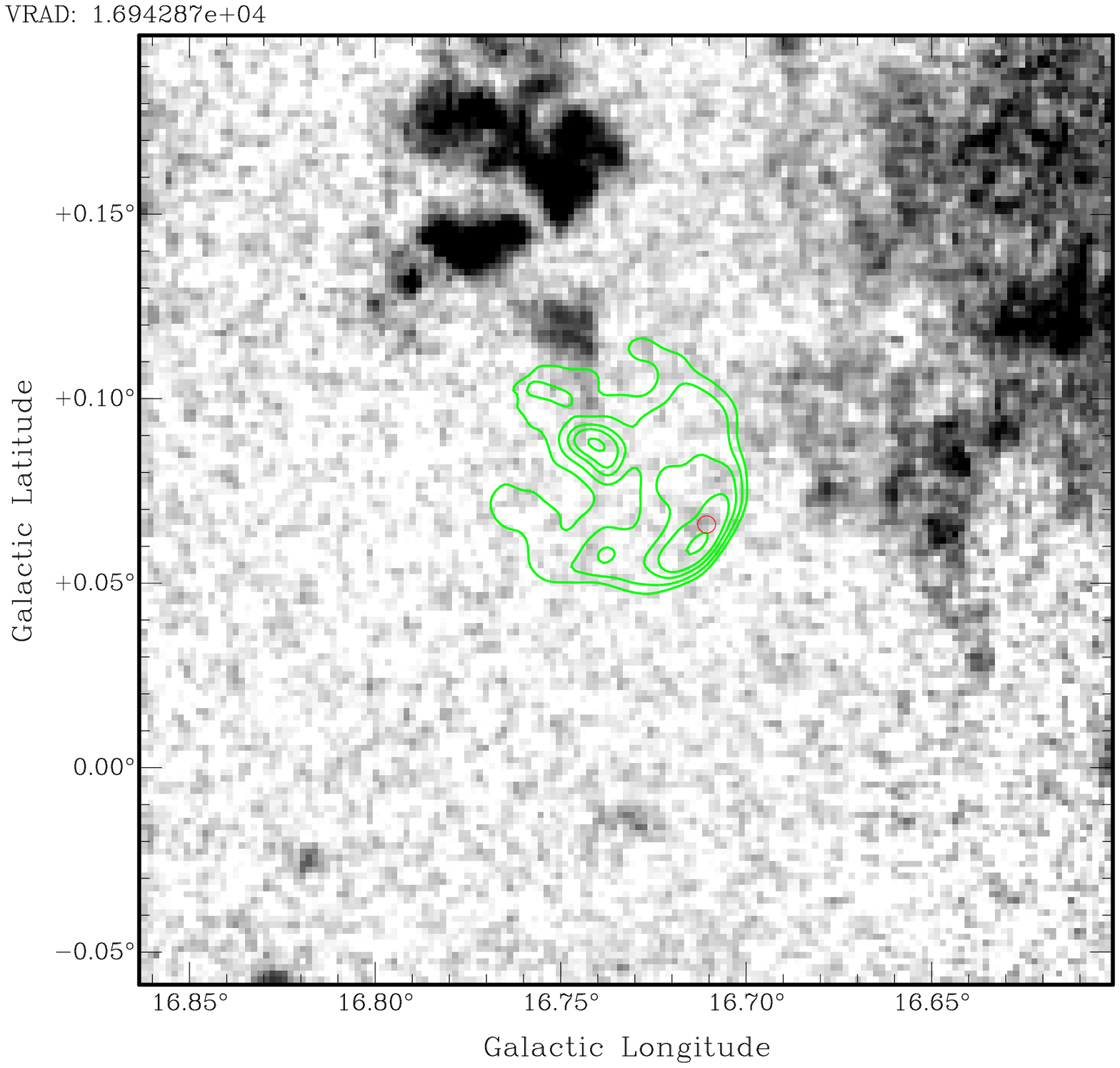}\includegraphics[width=0.33\textwidth]{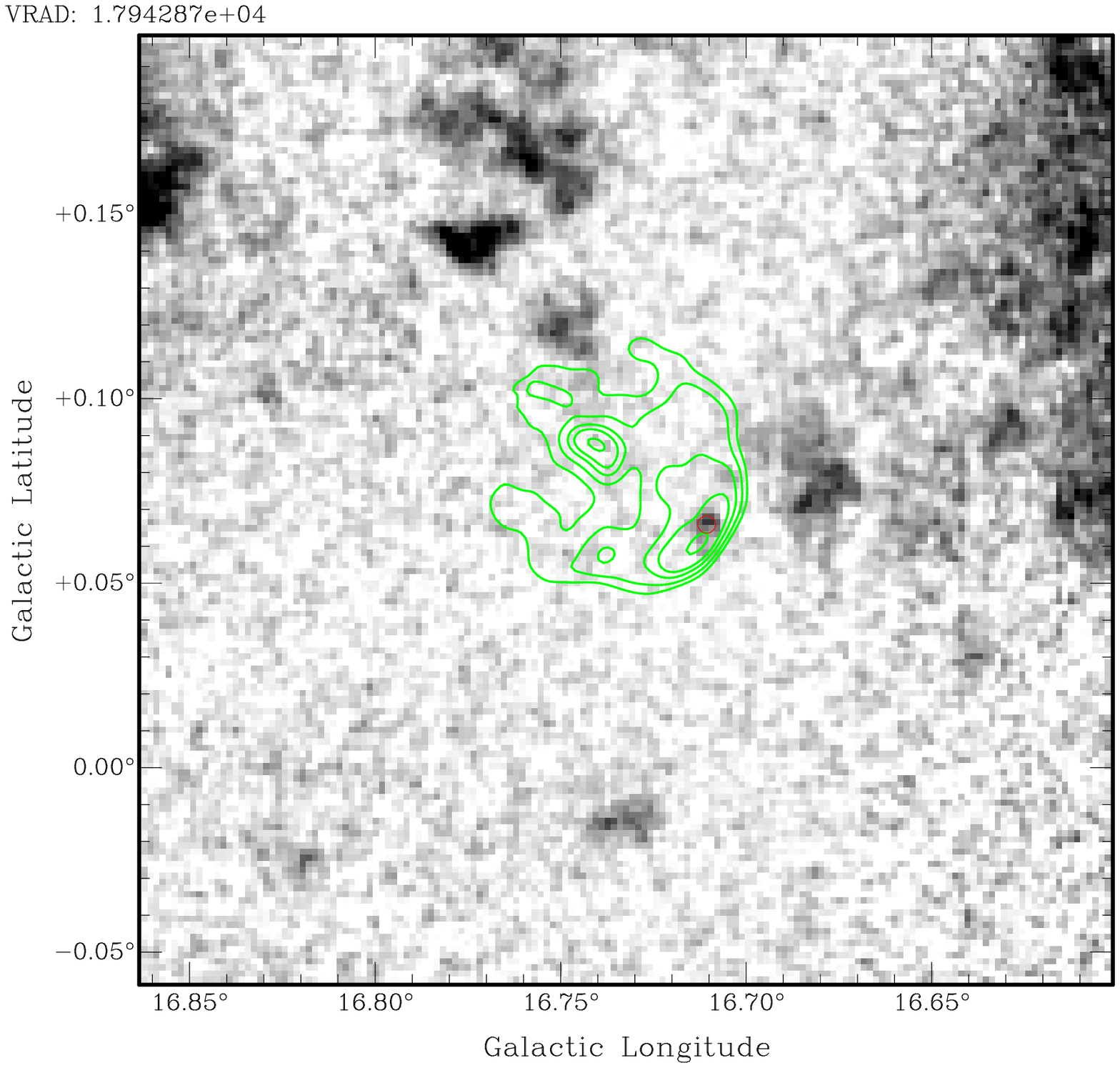}}
\centerline{\includegraphics[width=0.33\textwidth]{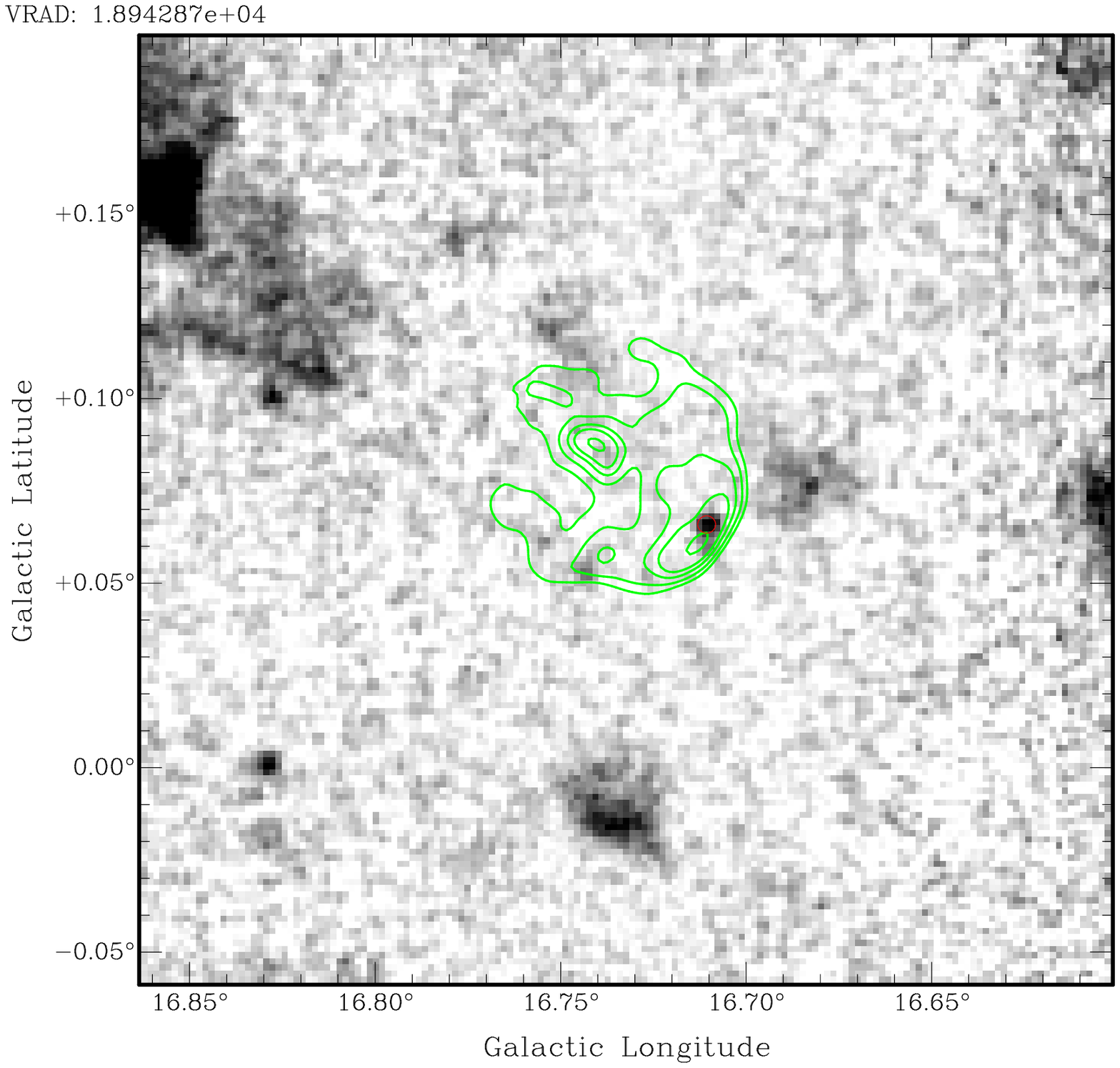}\includegraphics[width=0.33\textwidth]{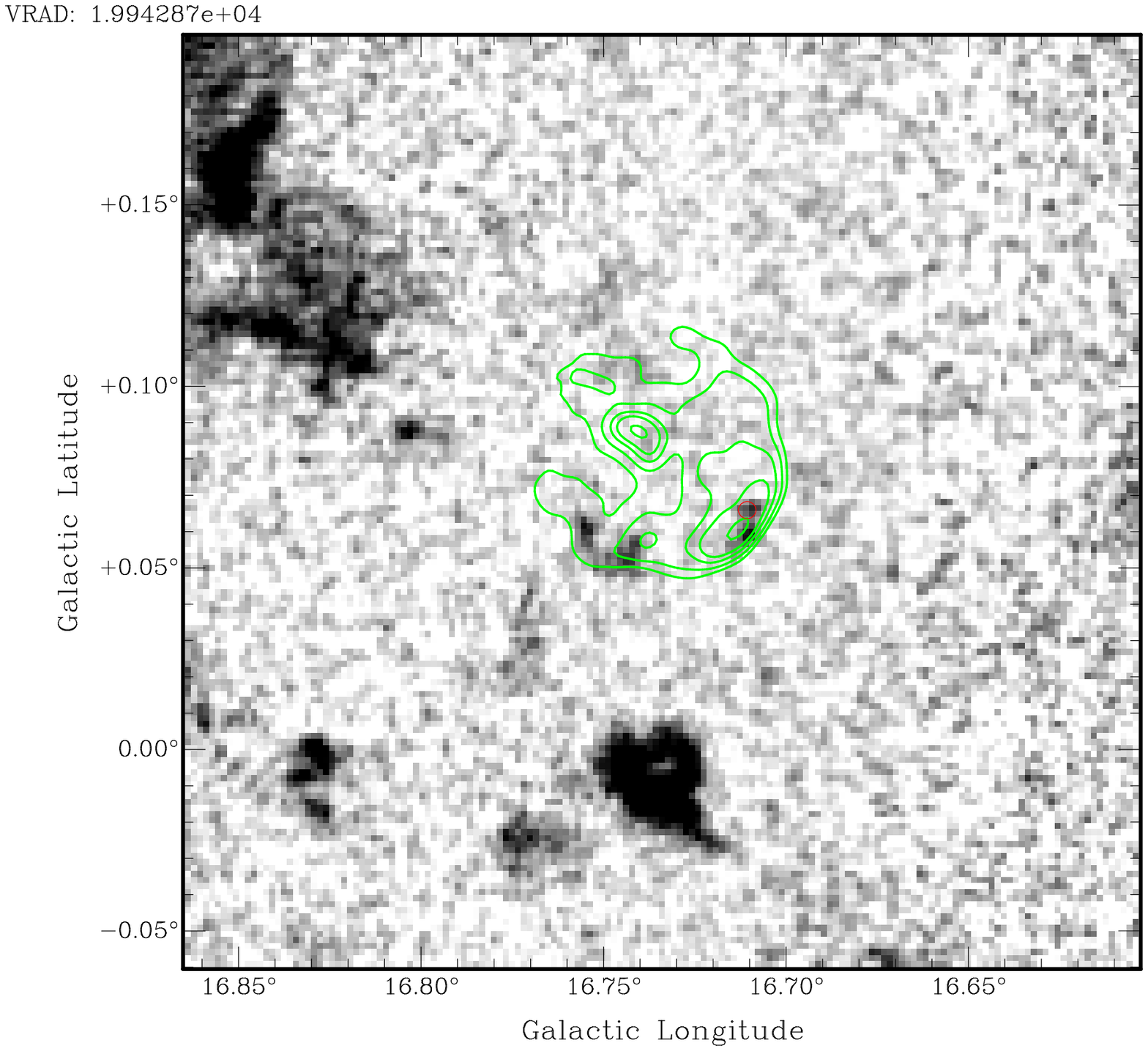}\includegraphics[width=0.33\textwidth]{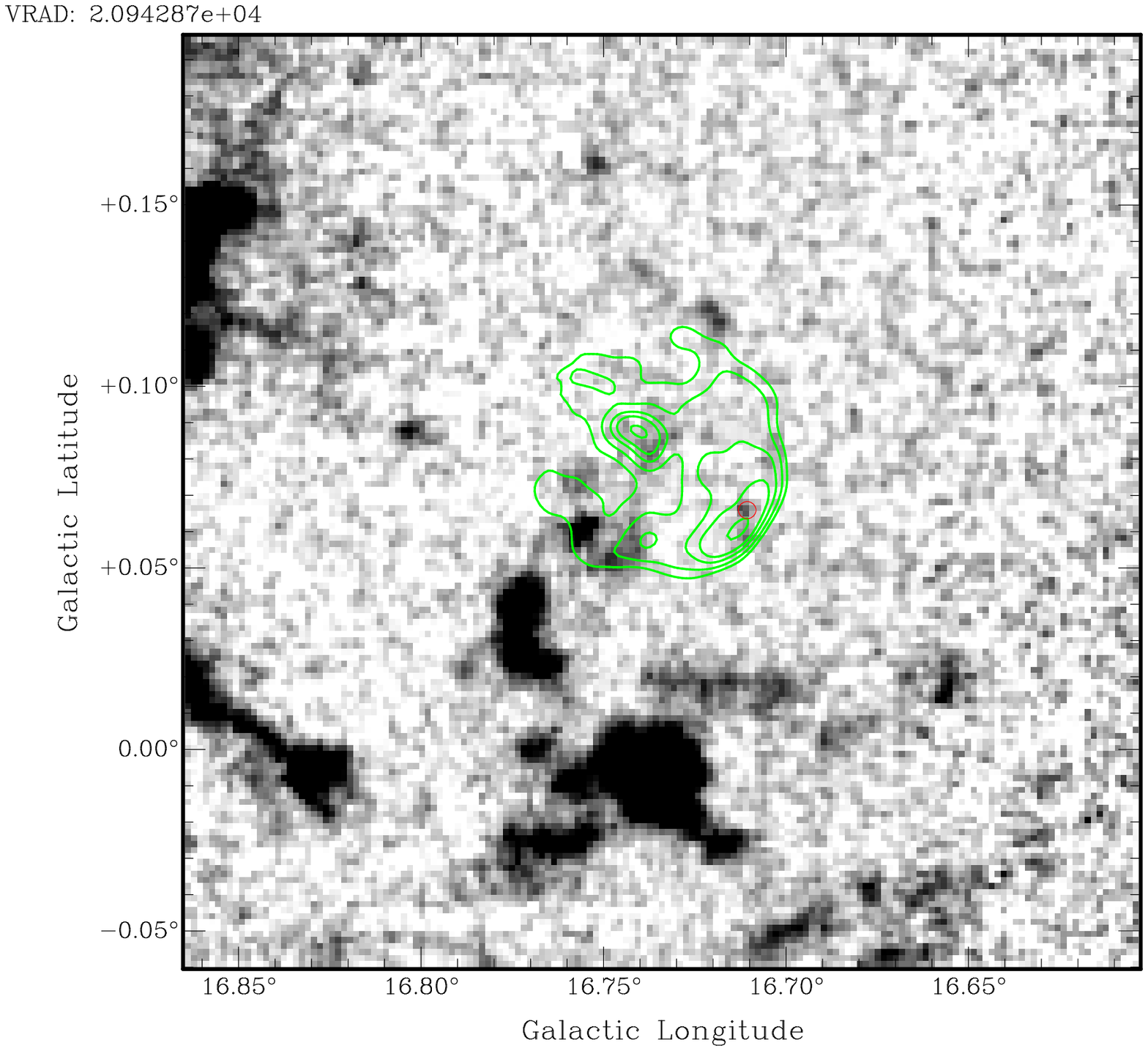}}
\centerline{\includegraphics[width=0.33\textwidth]{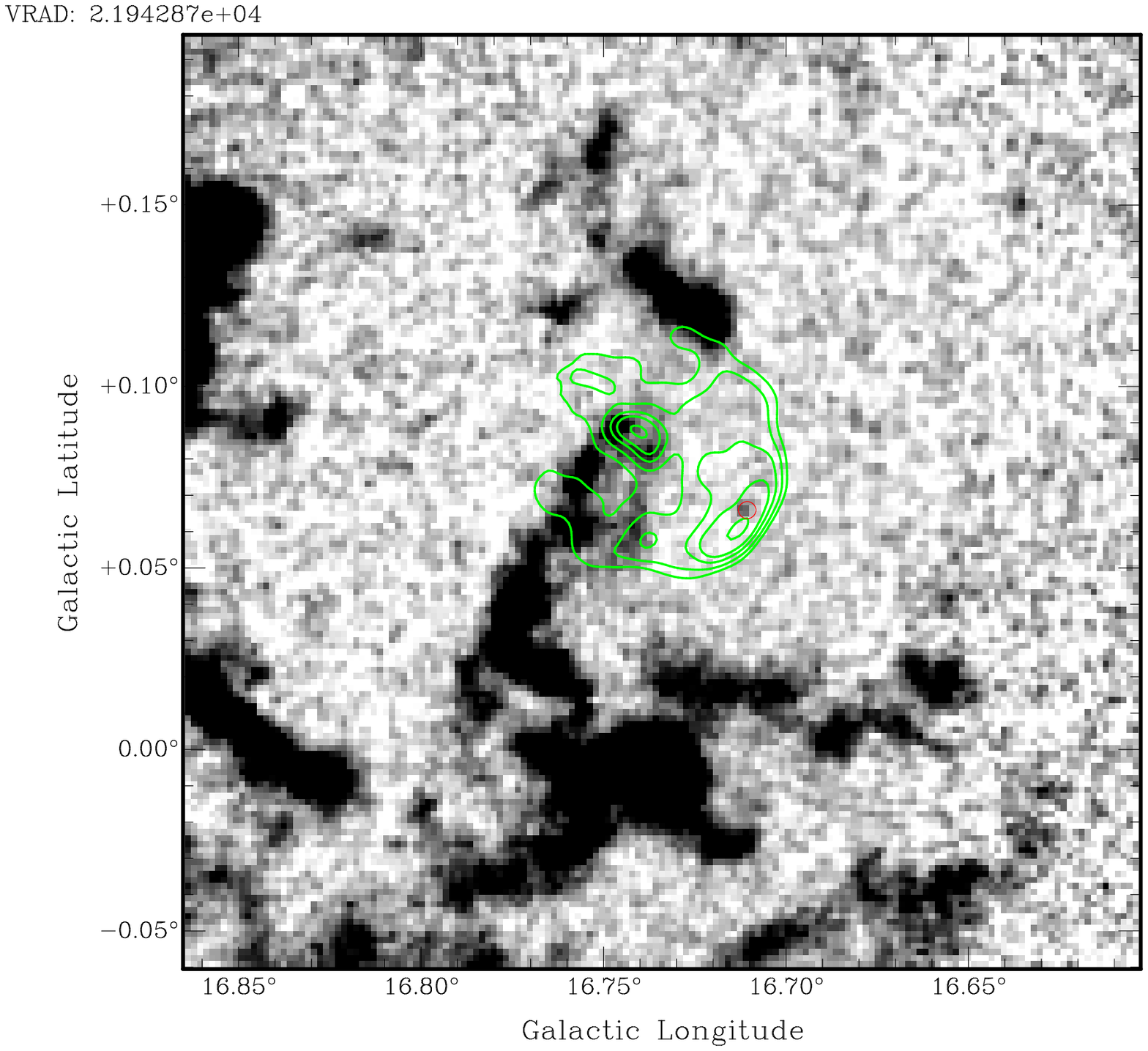}\includegraphics[width=0.33\textwidth]{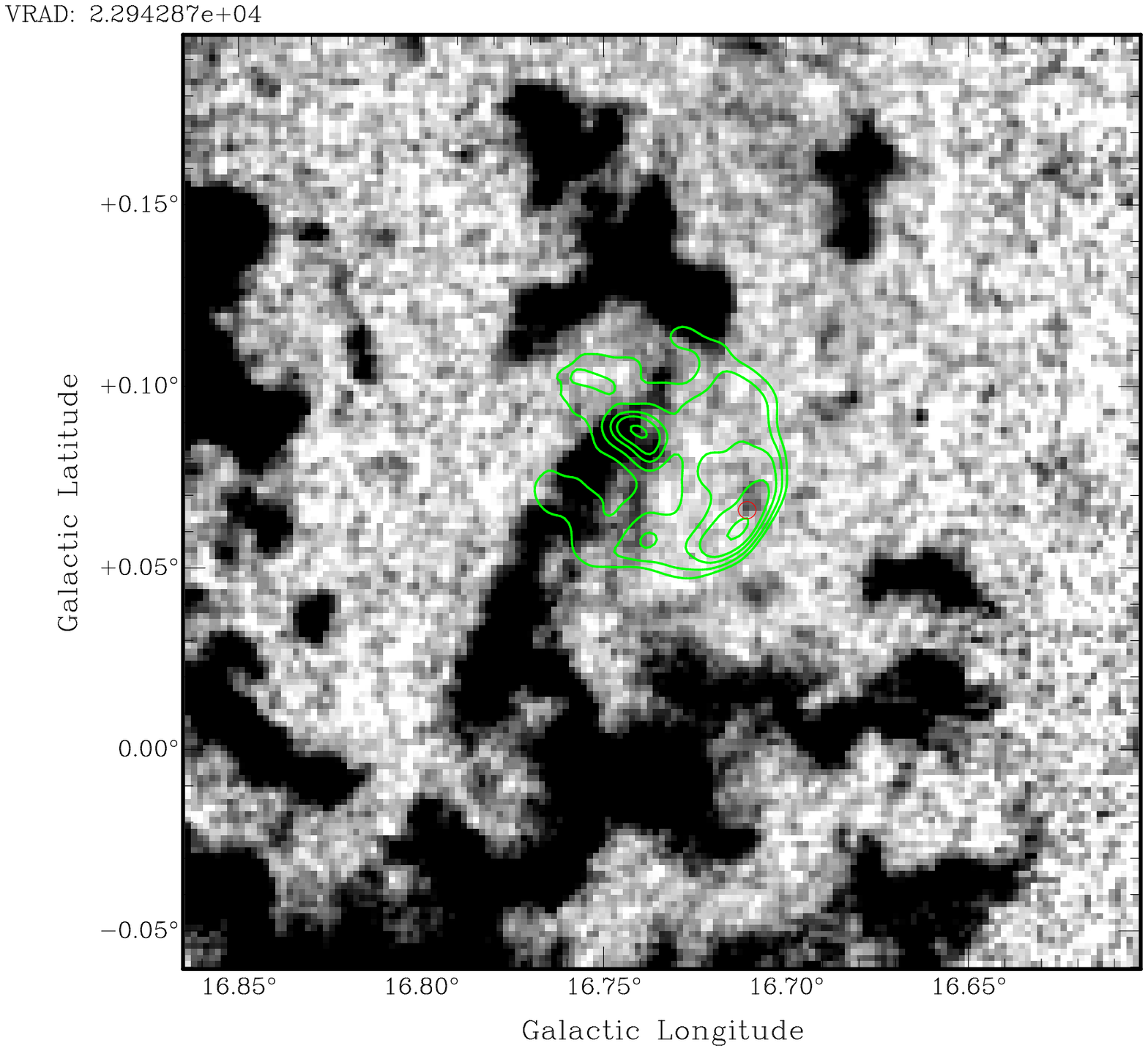}\includegraphics[width=0.33\textwidth]{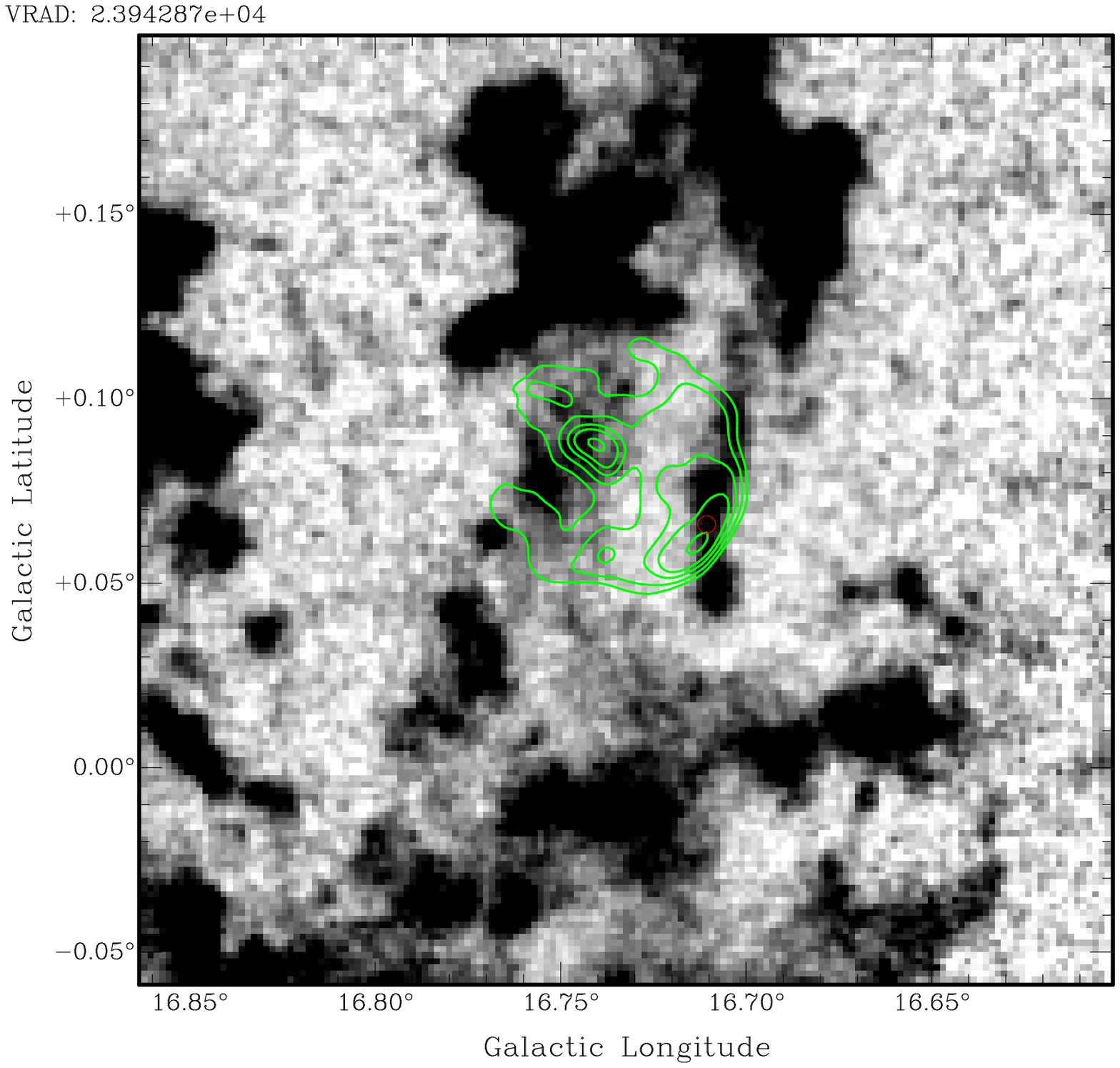}}
\centerline{\includegraphics[width=0.33\textwidth]{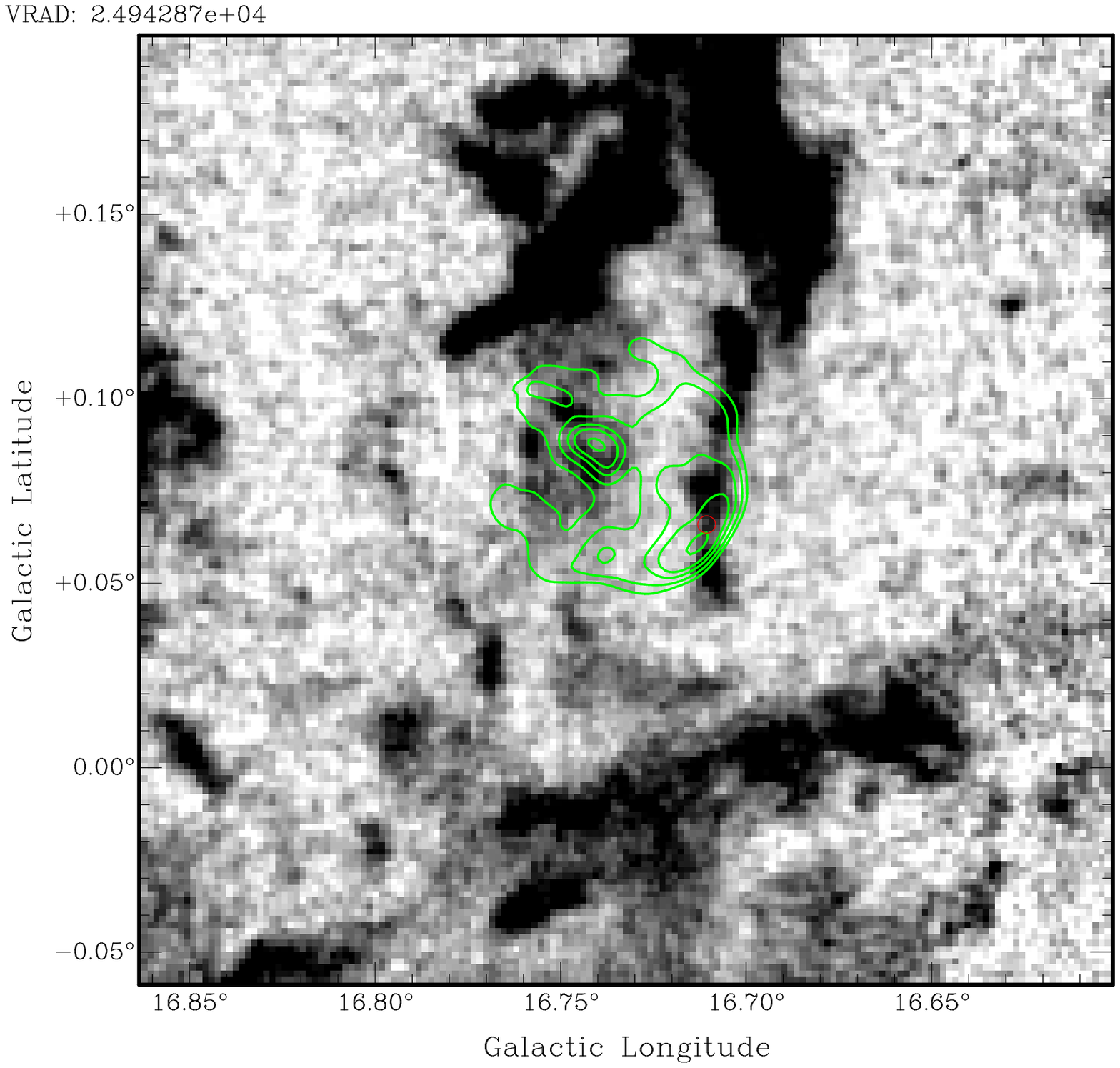}\includegraphics[width=0.33\textwidth]{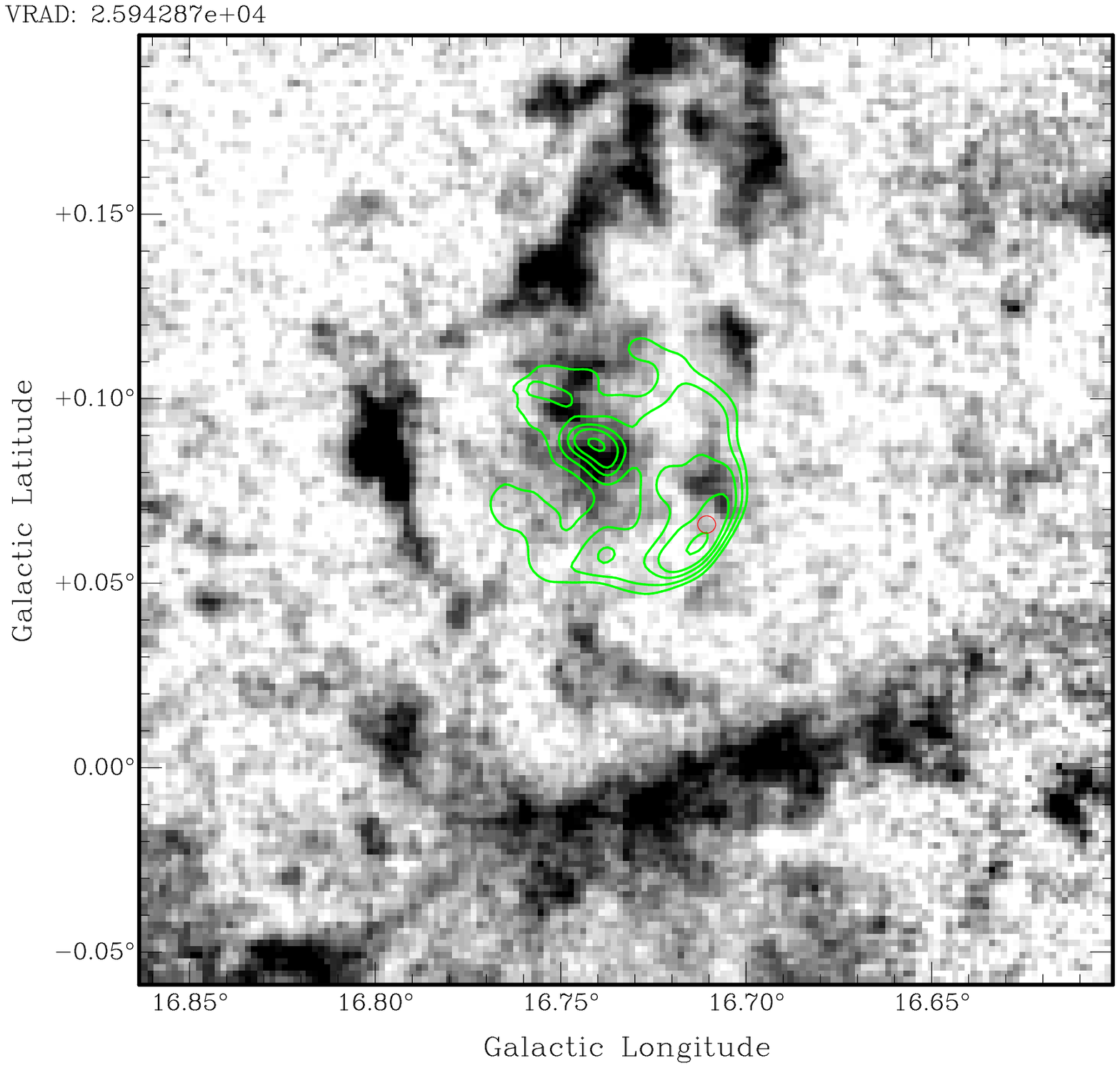}\includegraphics[width=0.33\textwidth]{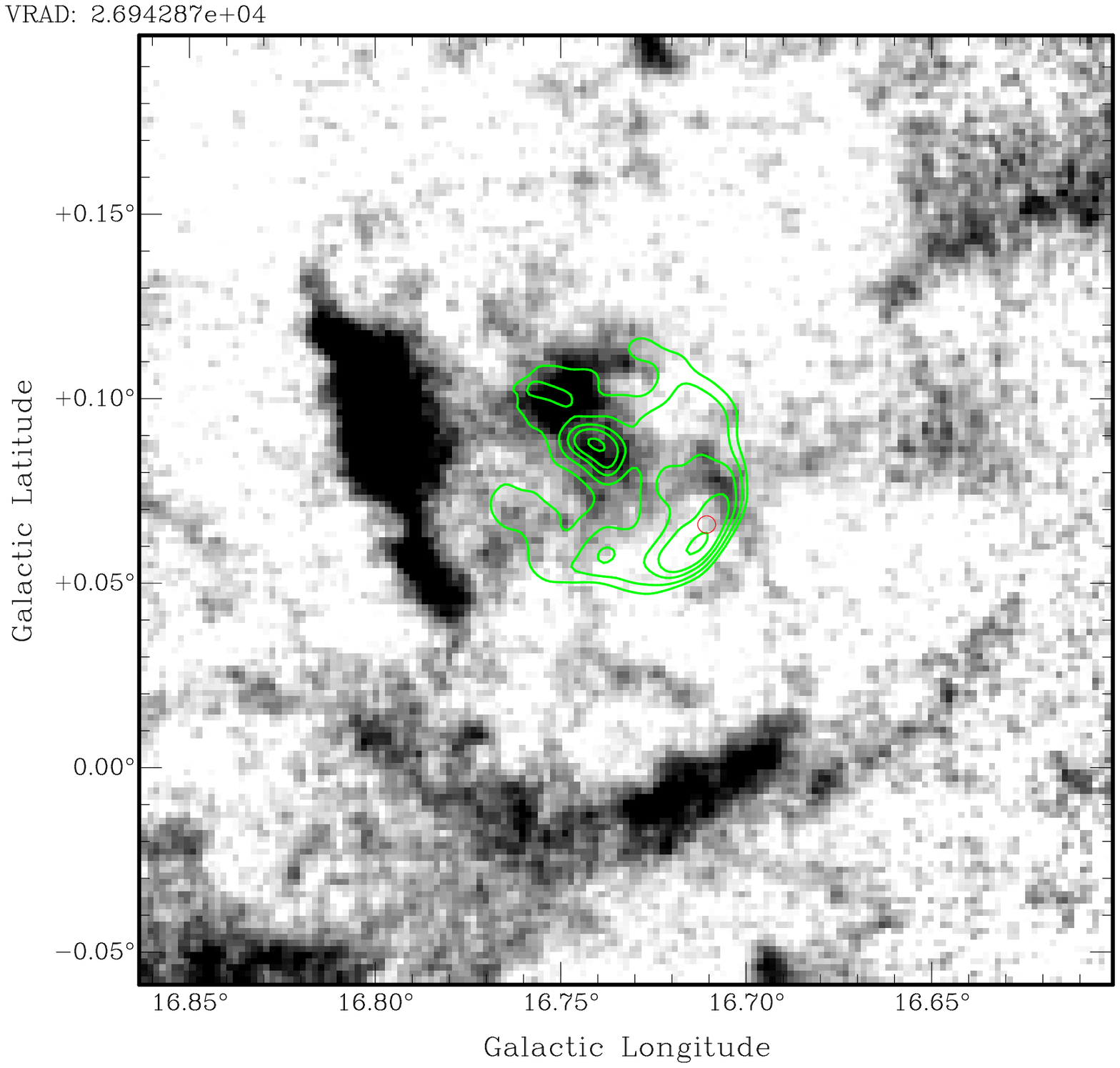}}
\caption{G16.7+0.1: The $^{12}$CO (J=3-2) channel maps from 16 \kms to 27 \kms. The contours are same as Fig.1. The position of the 1720 OH maser is labelled by the red circle.}
\label{fig:co}
\end{figure*}

We also create 12 channel-maps with the CO (J=3-2) emission data. The velocity ranges from 16 \kms to 27 \kms (Fig.~\ref{fig:co}). A bright small cloud at $\sim$19 \kms can be seen at the maser site (l=16.71, b=0.07). Weak CO (J=3-2) emissions also appear at the same site from 20 \kms to 23 \kms and gradually move into the 25 \kms southern cloud. This supports Reynoso \& Mangum¡¯s suggestion that the southern cloud was accelerated from 20 \kms to 25 \kms by the shock.\\

With the distance of 14 kpc, we estimate some basic parameters for G16.7+0.1, e.g. the X-ray luminosity, postshock density, X-ray-emitting mass and thermal energy, by assuming the plasma in the SNR shell is thermal \cite{Helfand2003} (see Table 2). Here we show the process of the calculation when $kT$ = 1 keV.
The electron density and proton density are also calculated by applying the equation L$_x$ = n$_e$n$_H\Lambda$(T)V, where L$_x$ is X-ray luminosity, n$_e$ is electron density, n$_H$ is proton density, $\Lambda$(T) is cooling function,
and V is the volume of material emitting X-rays.
We take n$_e$ = 1.2n$_H$ by assuming that hydrogen and helium inside the SNR have a cosmic abundance ([He]/[H] = 0.1) and are ionized.
$\Lambda$(T) is from the code CLOUDY (version 17.0) \cite{Ferland2017}.
To get the volume, we subtract the volume of central pulsar wind nebula (PWN) from the SNR.
The angular diameters of the whole SNR and the PWN are $4.5'$ and $1.5'$ respectively.
 We obtain a proton density of 0.12 cm$^{-3}$ for the shell, an X-ray-emitting mass of 10.7 M$_{\odot}$ and thermal energy of 9.2 $\times$ 10$^{48}$ erg.

\begin{table}
\begin{center}
\caption{Physical Parameters for Both SNRs}
\setlength{\tabcolsep}{2mm}
\begin{tabular}{ll}
\hline
\hline
 &  G16.7+0.1  \\
\hline
Distance:&  $\sim$ 14 kpc\\
Maximum HI absorption velocity:  & 138 km/s \\
Associated 1720 OH maser velocity:    &20 km/s \\
\hline
 &  G15.9+0.2  \\
 \hline
 Distance:&7-16 kpc\\
Maximum HI absorption velocity:  & 138 km/s \\
Likely associated CO emission velocity: & 20 km/s \\
\hline
\hline
\end{tabular}
\end{center}
\end{table}

\begin{table}
\begin{center}
\caption{Parameters for G16.7+0.1 at different temperatures}
\setlength{\tabcolsep}{1mm}
\begin{tabular}{lccccccc}
\hline
\hline
 kT  & L$_x$ &  SNR radius & PWN radius  & ${\Lambda}$(T) & n$_H$  & M$_{x}$ & E$_{thermal}$  \\
 (keV) & (10$^{34}$ erg s$^{-1}$) & (pc) & (pc) & (10$^{-23}$erg cm$^{-3}$s$^{-1}$) & (cm$^{-3}$) & (M$_{\bigodot}$) & (10$^{48}$ erg) \\
\hline
0.5 & 43 & 9.1 & 3.1 & 3 & 0.35 & 35 & 40\\
1.0 & 5.3 & 9.1 & 3.1 & 3.8 & 0.12 & 10.7 & 9.2\\
1.5 & 2.2 & 9.1 & 3.1 & 4.7 & 0.05 & 5 & 7.1\\
\hline
\hline
\end{tabular}
\end{center}
\end{table}
\subsection{SNR G15.9+0.2}
There are several previously studies on the distance of G15.9+0.2.
Caswell et al. (1982) \cite{Caswell1982} derived its distance of 16.7 kpc, and Pavlovic et al. (2014) \cite{Pavlovic2014}
obtained 10.4 or 18.4 kpc based on the $\Sigma$-D relation.
Reynolds et al. (2006) \cite{Reynolds2006} suggested 8.5 kpc for the remnant based on the large column density of 4 $\times$ 10$^{22}$ cm$^{-2}$ from the X-ray absorption.
Klochkov et al. (2016) estimated the distance of its central point source CXOUJ181852.0-150213  to be 10-30 kpc via the carbon atmosphere model fitting \cite{Klochkov2016}.
Maggi \& Acero (2017) \cite{Maggi2017} found a lower limit distance of 5 kpc based on the CO/HI velocity $<$ 60 \kms. Our HI method shows an independent measurement and a better constraint for the remnant's distance (7 to 16 kpc) (see Table 1).

\begin{figure}
    \centering
    \includegraphics[width=0.8\textwidth]{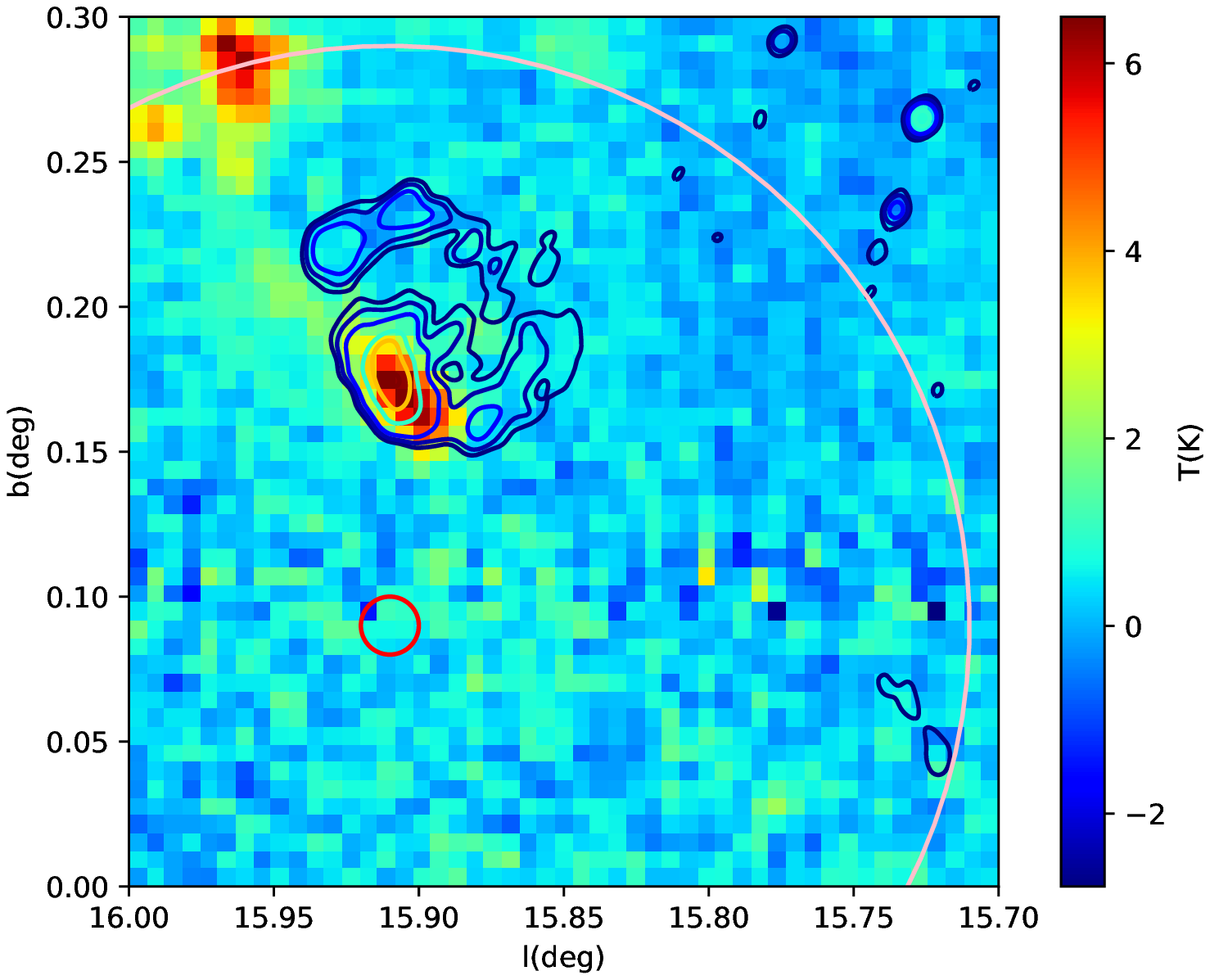}
    \caption{G15.9+0.2:The total \co channel map integrated from 20.76 to 21.83 \kms. The contour is the 1.4 GHz continuum
    map. The red circle is the central coordinate (l=15.91,b=0.09) \cite{Abeysekara2017} of 2HWC J1819-150*,
    and the pink circle shows the beam size of the highest angular resolution of 0.2$^{\circ}$ for \textit{HAWC}.}
\label{fig:contour}
\end{figure}

Recently, HAWC (High Altitude Water Cherenkov Observatory)
team released the new 2HWC catalog (Abeysekara
et al. 2017). This contains a TeV source, 2HWC
J1819-150*, overlapping with SNR G15.9+0.2. The positional uncertainty of TeV source 2HWC J1819-150* is consistent with G15.9+0.2
(see Fig.~\ref{fig:contour}).
Since more and more evidence from both observations and theories reveal that SNR-MC interactions can generate TeV
emission, here arises the queston: Is the 2HWC J1819-150* generated by the interaction between SNR G15.9+0.2 and an MC?

The known TeV source, HESS J1818-154 is $0.5^{\circ}$ away from the center of 2HWC J1819-150* and has been suggested to be
associated with SNR G15.4+0.1.
Their angular distance is larger than the highest resolution of \textit{HAWC},
$0.2^{\circ}$, and the angular distance between 2HWC J1819-150* and G15.9+0.2 is less than $0.1^{\circ}$.
In addition, the $\gamma$-ray fluxes of 2HWC J1819-150* and HESS J1818-154 at 7 TeV are 59.0$^{+7.9}_{-7.9}$ and
4.5$^{+7.9}_{-3.2}$ $\times$10$^{15}$ TeV$^{-1}$cm$^{-2}$s$^{-1}$ respectively \cite{Abeysekara2017, Abramowski2014},
which implies they are different sources.
In conclusion, 2HWC J1819-150* is nearer to G15.9+0.2 than G15.4+0.1 and likely not related to HESS J1818-154.
The fact that HAWC does not detect HESS J1818-154 is probably due to its lower sensitivity than H.E.S.S
or the different energy responses between H.E.S.S and HAWC.
However, H.E.S.S did not detect 2HWC J1819-150*, which possibly can be explained if 2HWC J1819-150* is a variable source.

A bright \co filament at about 20 \kms is located on the eastern shell of G15.9+0.2 (see Fig.6),
so, there is a high possibility that the eastern part of G15.9+0.2 is interacting with the CO cloud. If so, the distance of G15.9+0.2 is 14 kpc, consistent with our HI distance measure.

\section{Summary}
HI absorption spectrum is an independent way for the distance estimate of Galactic radio sources.
We obtain a distance of about 14 kpc for SNR G16.7+0.1 by analyzing the 1720 MHz OH maser and the HI absorption spectra towards this remnant. We give a reliable distance constraint (7 to 16 kpc) for G15.9+0.2 from HI absorption, and possibly 14 kpc if the remnant is associated with the CO cloud at 20 \kms.  Based on our distance measurements, we discussed other physical parameters of the two SNRs.

\subsection{Acknowledgments}

We acknowledge support from the National Key R\&D Programs of China (2018YFA0404203,2018YFA0404202) and NSFC programs (11603039, U1831128, U1738125). D.A.Leahy was supported by the Natural Science and Engineering Research Council of Canada.
This publication makes use of molecular line data from the Boston University-FCRAO Galactic Ring Survey (GRS).

\section{Reference}

\bibliographystyle{iopart-num}
\bibliography{mydb}

\end{document}